\documentclass[twocolumn,preprintnumbers,showpacs,aps,prl,floats,amssymb]{revtex4}
\usepackage{graphicx}
\begin{document}
\author{Matthias Troyer$^{(1,2)}$}
\author{Stefan Wessel$^{(1)}$}
\author{Fabien Alet$^{(3,1,2)}$}
\affiliation{$^{(1)}$Theoretische Physik, ETH Z\"urich, CH-8093 Z\"urich,
  Switzerland}
\affiliation{$^{(2)}$Computation Laboratory, ETH Z\"urich, CH-8092 Z\"urich,
  Switzerland}
\affiliation{$^{(3)}$Laboratoire de Physique Quantique, Universit\'e Paul
  Sabatier, 31062 Toulouse, France}
\date{\today}
\title{Wang-Landau sampling for quantum systems: \\ 
algorithms to overcome tunneling problems and calculate the free energy}
\begin{abstract}
We present a generalization of the classical Wang-Landau algorithm
[Phys. Rev. Lett. {\bf 86}, 2050 (2001)] to quantum systems. The algorithm
proceeds by stochastically evaluating the coefficients of a high temperature
series expansion or a finite temperature perturbation expansion to arbitrary
order. Similar to their classical counterpart, the algorithms are efficient at
thermal and quantum phase transitions, greatly reducing the tunneling problem
at first order phase transitions, and allow the direct calculation of the free
energy and entropy.
\end{abstract}
\pacs{02.70.Ss,05.10.Ln,05.30.-d,75.40.Mg}
\maketitle

%%%

%\section{Introduction}

Monte Carlo simulations in statistical physics now have a history of nearly
half a century starting with the seminal work of Metropolis
\cite{metropolis}. While the Metropolis algorithm has been established as the
standard algorithm for importance sampling it suffers from two problems: the
inability to directly calculate the partition function, free energy or
entropy, and critical slowing down near phase transitions and in disordered
systems. 

In a standard Monte Carlo algorithm a series of configurations is generated
according to a given distribution, usually the Boltzmann distribution in
classical simulations. While this allows the calculation of thermal averages,
it does not give the partition function, nor the free energy. They can only be
obtained with limited accuracy as a temperature integral of the specific heat,
or by using maximum entropy methods \cite{maxentcv}. 

The problem of critical slowing down has been overcome for second order phase
transitions by cluster update schemes for classical \cite{swendsenwang} and
quantum systems \cite{loop,wiese,sseloops,worm}. For first order phase
transitions and systems with rough free energy landscapes a decisive
improvement was achieved recently by a new algorithm for classical systems due
to Wang and Landau \cite{wanglandau}. In contrast to related methods -- such
as the multicanonical \cite{multicanonical} or the broad histogram
\cite{broadhistogram} method -- this new algorithm scales to large systems,
does not suffer from systematic errors and needs no a priori knowledge.
The key idea is to calculate the density of states $\rho(E)$ directly by a
random walk in energy space instead of performing a canonical simulation at a
fixed temperature. By visiting each energy level $E$ with a probability
$1/\rho(E)$, this algorithm achieves a flat histogram and good precision  over
the whole energy range.
Besides being efficient at first and second order phase transitions this
algorithm allows the direct calculation of the free energy from the partition
function $Z=\sum_E \rho(E)e^{-E/k_B T}$. The internal energy, entropy,
specific heat and other thermal properties are easily obtained as well,
by differentiating the free energy. Within a year of publication this
algorithm has been improved using $N$-fold way \cite{nfold} and multibondic
\cite{multibondic} sampling schemes, has been applied to Potts models
\cite{potts} and generalized to reaction coordinates \cite{reaction},
continuum models \cite{continuum}, polymer films \cite{polymer}, and to
protein folding \cite{proteins}.

Since simulations of quantum systems suffer from the same problems as
classical simulations, in particular from long tunneling times at first order
phase transitions and the inability to calculate the free energy directly, an
extension of this algorithm to quantum systems is highly desired.
Here we present two such algorithms. The first one is based upon a high
temperature series expansion. Similarly to the classical algorithm it allows
the calculation of the free energy as a function of temperature, making it
ideal for the study of thermal phase transitions. The second algorithm renders
the high temperature series expansion into a perturbation expansion in one of
the coupling constants and is suitable for the investigation of quantum phase
transitions.

Quantum Monte Carlo algorithms start by mapping the quantum system
to a classical system. This can be done either through a discrete
\cite{suzukitrotter} or continuous time \cite{wiese} path integral
representation or by a stochastic series expansion (SSE) in the inverse
temperature \cite{sse}. While our algorithms can be formulated in both
representations, we here present the SSE version as it is the easiest and most
natural representation for most problems.

%\section{High Temperature Expansion}

We start by expressing the partition function as a high temperature expansion
\begin{equation}
Z={\rm Tr}e^{-\beta H}=\sum_{n=0}^\infty \frac{\beta^n}{n!}{\rm
  Tr}(-H)^n\equiv \sum_{n=0}^\infty g(n)\beta^n,
\label{eq:zquantum}
\end{equation}
where the $n$-th order series coefficient $g(n)={\rm Tr}(-H)^n/n!$  will play
the role of the density of states in the classical algorithm. 
The algorithm performs a random walk in
the space of series expansion coefficients, achieves a flat histogram in their
orders $n$ and calculates the coefficients $g(n)$.
Employing Eq. (\ref{eq:zquantum}) we can then calculate the free
energy, internal energy, entropy and specific heat directly. 
Thermal averages of observables can be measured as in conventional Monte Carlo
algorithms by recording a separate histogram for the expectation values at
each order.

Next we note that in a computer simulation the series expansion
(\ref{eq:zquantum}) needs to be truncated at an order $\Lambda$. Since the
orders relevant for a given inverse temperature $\beta$ are sharply peaked
around $\beta |E(\beta)|$, where $E(\beta)$ is the mean energy at inverse
temperature $\beta$, this cutoff does not introduce a systematic error.
Its main consequence is to restrict the accessible temperature range to 
$\beta \lesssim \Lambda/E(\beta)$.

The next step is to introduce a complete set of basis states
$\{|\alpha\rangle\}$, and to decompose the Hamiltonian $H$ into diagonal and
offdiagonal bond terms $H_b^{(a)}$. For simplicity we restrict the following
discussion to a spin-$1/2$ Heisenberg model where this decomposition for a
bond $b=(i,j)$ reads
$H_{(i,j)}^{(d)}=JS_i^zS_j^z-J/4$, and
$H_{(i,j)}^{(o)}=J/2(S_i^+S_j^-+S_i^-S_j^+)$.
The offset $-J/4$ is added to the diagonal part in order to render the matrix
elements of $-H$ nonnegative. 
Using this decomposition, we can write the partition function as
\cite{sseloops,sse}
\begin{equation}
  Z = \sum_\alpha \sum_{\{S_\Lambda\}} \frac{\beta^n (\Lambda-n)!}{\Lambda!}
   \langle\alpha| \prod_{i=0}^\Lambda (-H^{(a_i)}_{b_i}) |\alpha\rangle,
\label{eq:ssel}
\end{equation}
where the operator string $S_\Lambda=((b_1,a_1),\ldots,(b_\Lambda,a_\Lambda))$
is a concatenation of $n$ bond operators and $\Lambda-n$ unit operators.

Comparing Eq. (\ref{eq:zquantum}) to Eq. (\ref{eq:ssel}) we see
that we can obtain $g(n)$ by counting the number of times a configuration with
$n$ non-unit operators is observed during a simulation at an inverse
temperature $\beta=1$. As the dynamic range of $g(n)$ spans thousands of
orders of magnitude [$g(0)/g(\Lambda)$ is up to $10^{10000}$ for the examples
given below] simply collecting a histogram will not be effective. Therefore a
variant of the classical Wang-Landau method \cite{wanglandau} will be
employed: by reweighting a configuration of $n$-th order with $1/g(n)$ a flat
histogram of the orders $n$ can be achieved, thus sampling all orders equally
well.

Since $g(n)$ is initially unknown we start with the (bad) guess $g(n)=1$. Each
time a configuration of $n$-th order is visited, $g(n)$ is multiplied by a
factor $f$, i.e. $g(n)\leftarrow f g(n)$. In our implementation we store the
logarithms of these quantities to avoid overflow problems. The random walk is
performed until the histogram $H(n)$ -- counting the number of times a
configuration with $n$ operators is observed -- is reasonably flat. Similar to
the classical case a maximum deviation of 20\% from the mean value turned out
to be reasonable. The multiplicative increase of $g(n)$ is essential for the
success of the algorithm. Only that way the large range of $g(n)$ can be
mapped out in reasonable time, and $g(n)$ converges rapidly to a rough
estimate of the true distribution.
Once the histogram is flat, it is reset to zero, $f$ is decreased, in our case
by $f\leftarrow\sqrt{f}$, and the process starts again, refining $g(n)$
further with smaller steps. This procedure is repeated until $f$ gets as
small as $\exp(10^{-8})$, so that an accurate estimate of $g(n)$ with only
negligible systematic errors will be available. The accuracy of the free
energy and other calculated quantities is given by the statistical error
which, as usual, scales with $1/\sqrt{N_{\rm MC}}$ where $N_{\rm MC}$
is the number of Monte Carlo steps.
The overall normalization of $g(n)$ follows from $g(0)$ being equal to the
dimension of the Hilbert space, which for a spin-$1/2$ Heisenberg model on a
lattice with $N$  sites is $2^N$.
The initial choice of $f$ is very important. Too small starting values result
in  long computation times, while too large values give extreme fluctuations
in the initial iterations. As in the  classical
Wang-Landau method a good choice is to let $f^{N_{\rm MC}}$ be of the same
order of magnitude as the total number of configurations, which is of order
$2^NN^\Lambda$. Since usually $\Lambda\gg N$, a good initial value is $\ln f
\approx (\Lambda\ln N)/N_{\rm MC}$. 

\begin{figure}
\includegraphics[width=7cm]{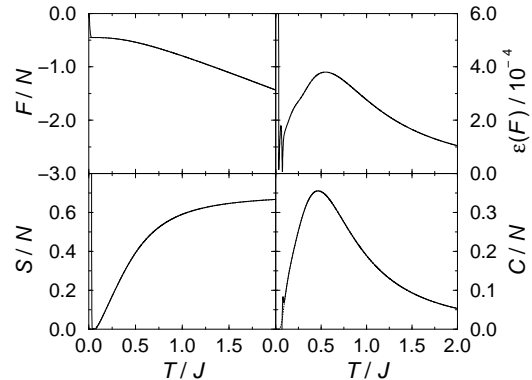}
\caption{Free energy $F$ , entropy $S$  and specific heat $C$ of an $N=10$
  site antiferromagnetic Heisenberg chain. Solid lines correspond to the MC
  results, indistinguishable from the dotted lines for the exact results. Also
  shown is the relative error $\varepsilon(F$) of $F$ compared to the exact
  result.}
\label{fig:chain}
\end{figure}

To finish the description of the algorithm we discuss the update steps in more
detail. Any of the known update algorithms, employing local \cite{sse}, or
cluster \cite{sseloops} updates can be used. The only change
in the acceptance probabilities from standard SSE algorithms is to set
$\beta=1$ and to include an additional factor $g(n)/g(n')$ in the acceptance
probability for any move that changes the number of operators from $n$ to
$n'$.
As an example we discuss the Heisenberg antiferromagnet. There the optimal
algorithm is the loop algorithm, which in the SSE representation consists of
two parts: diagonal updates and loop updates. In a diagonal update step the
operator string positions $l=1,...,\Lambda$ are traversed in ascending order. 
Empty and diagonal operators can be exchanged with each another.
Using the notation
$
  |\alpha(l)\rangle = \prod_{i=1}^l H_{b_i}^{(a_i)} |\alpha\rangle
$
for the state obtained by acting on $|\alpha\rangle$ with the first $l$ bond
operators, and $M$ for the total number of interacting bonds on the 
lattice, the update probabilities are
\begin{eqnarray}
P[H^{(0)}{(l)}\rightarrow H^{(d)}_b{(l)}] &=&
\frac{g(n)M\langle\alpha(l)|H^{(d)}_b
  |\alpha(l)\rangle}{g(n+1)(\Lambda-n)},\label{eq:update} \\
P[H^{(d)}_b{(l)}\rightarrow H^{(0)}{(l)}] &=&
\frac{g(n)(\Lambda-n+1)}{g(n-1)M\langle\alpha(l)|H^{(d)}_b
  |\alpha(l)\rangle}\nonumber,
\end{eqnarray}
where $P>1$ is interpreted as $P=1$.
This choice of update probabilities requires the least changes to an existing
SSE program. Alternatively the factors $M$, $\Lambda-n$ and $\Lambda-n+1$ can
be dropped from the update probabilities, thus simplifying the algorithm. To
correct for this omission, the obtained $g(n)$ must then be multiplied by
$M^n (\Lambda-n)!/\Lambda!$.
At each level $l$, independent of whether an update was performed or not
(e.g. when the operator is off-diagonal) $g(n)$ for the current value of $n$
is incremented by $f$. The second part of the update cycle, the loop update,
changes diagonal to off-diagonal bond operators without changing $n$ and
can be performed as in standard SSE algorithms. We refer to
Ref. \cite{sseloops} for details.

\begin{figure}
\includegraphics[width=7cm]{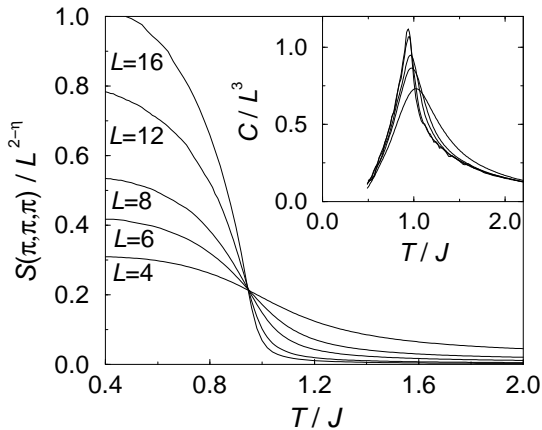}
\caption{Scaling plot of the staggered structure factor of a cubic
  antiferromagnet as a function of temperature, obtained from simulations at 
  a fixed temperature for various lattice sizes. The inset shows the
  specific heat as a function of temperature. The cutoff $\Lambda=500 (L/4)^3$
  restricts the accessible temperature range to $T\gtrsim 0.4J$.}
\label{fig:af}
\end{figure}

As a first example we show in Fig. \ref{fig:chain} results of calculations for
the free energy $F$, entropy $S$ and specific heat $C$ of an $N=10$ site
antiferromagnetic Heisenberg chain, and compare to exact results. Using $10^8$
sweeps, which can be performed in less than five hours on an 800 MHz
Pentium-III CPU, the errors can be reduced down to the order of $10^{-4}$. The
cutoff was set to $\Lambda=250$, restricting the accessible temperatures to
$T\gtrsim 0.05J$. The sudden departure of the Monte Carlo data from the exact
values below this temperature clearly shows this limit, which can be pushed
lower by increasing $\Lambda$. The sudden deviation becomes even more
pronounced in larger systems and provides a reliable indication for the range
of validity of the results. 

To illustrate the efficiency of the algorithm close to a {\it thermal
second order phase transition}, we consider the
Heisenberg antiferromagnet on a simple cubic lattice. From 
simulations of systems with $L^3$ sites,
$L=4,6,8,12,16$, we can calculate the staggered structure factor
$S(\pi,\pi,\pi)$ for any value of the temperature using the measured
histograms. Figure \ref{fig:af} shows the scaling plot of
$S(\pi,\pi,\pi)/L^{2-\eta}$ with $\eta=0.034$. The estimate for the critical
temperature $T_c=0.947 J$, obtained in only a couple of days on an 800 MHz
Pentium-III CPU, compares well with earlier estimates \cite{3dafm}. 

\begin{figure}
\includegraphics[width=7cm]{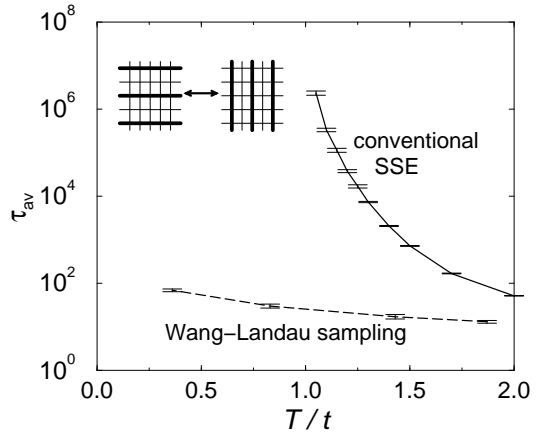}
\caption{Average tunneling times (in units of Monte Carlo sweeps) between
  horizontal and vertical arrangement of stripes in a hard core boson model at
  a ratio $V_2/t=3$ of next nearest neighbor repulsion to kinetic energy on a
  $8 \times 8$ sites lattice. The solid line is obtained using the standard
  SSE algorithm with directed loop updates. The dashed line is obtained using our
  algorithm, where the temperature is defined as the lowest temperature accessible in the simulation.}
\label{fig:tau}
\end{figure}

Next we demonstrate the efficiency of the algorithm at a {\it first order phase transition} by considering two-dimensional hardcore bosons with next-nearest neighbour
interactions \cite{hcb}. At low temperature and half filling this model is in an insulating phase with striped charge order and provides the simplest quantum mechanical model with stripes. We are currently investigating the thermal and quantum melting transitions of this stripe phase and probe for the existence of a nematic phase \cite{athens}. Simulations with conventional update schemes suffer from exponentially increasing
tunneling times needed to change the stripe orientation from a horizontal to a vertical arrangement. The flat histogram in the order $n$ in our algorithm reduces the
tunneling times by many orders of magnitude  already on small lattices (c.f.
Fig. \ref{fig:tau}) which demonstrates the efficiency of our algorithm at first order phase transitions.

%\section{Coupling expansion}

We now turn to our second algorithm, which applies to quantum phase
transitions. Instead of scanning a temperature range we vary one of the interactions at fixed temperature. Defining the Hamiltonian as $H=H_0+\lambda V$ we
rewrite the partition function Eq. (\ref{eq:zquantum}) as
\begin{equation}\label{eq:zlambda}
Z=\sum_{n=0}^\infty \frac{\beta^n}{n!}{\rm Tr}(-H_0-\lambda V)^n\equiv \sum_{n_{\lambda}=0}^\infty \tilde{g}(n_\lambda)\lambda^{n_\lambda},
\end{equation}
where on the right hand side we have collected all terms associated with $\lambda^{n_\lambda}$ into 
$\tilde{g}(n_\lambda)$. A similar algorithm can now be devised for this
perturbation expansion up to arbitrary orders by setting $\lambda=1$, replacing $M$ by $\beta M$
and $g(n)$ by $\tilde{g}(n_\lambda)$ in  Eq. (\ref{eq:update}). 
To normalize  $\tilde{g}(n_\lambda)$ there are two options. If $H_0$
can be solved exactly, $\tilde{g}(0)$ can be determined
directly. Otherwise, the normalization can be fixed using the first algorithm
to calculate $Z(\beta)$ at any fixed
value of $\lambda$.
Finally, even without normalization we can still obtain entropy and
energy differences.

\begin{figure}
\includegraphics[width=7cm]{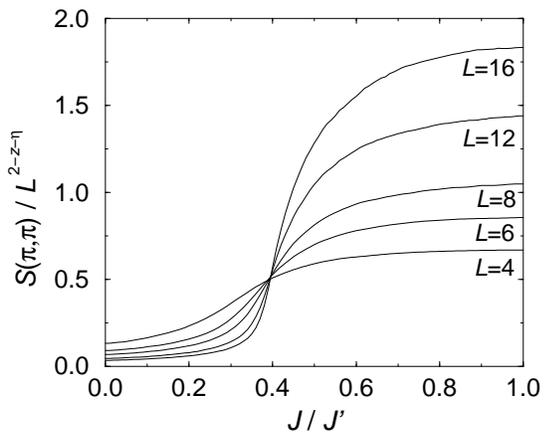}
\caption{Scaling plot of the staggered structure factor of a Heisenberg
  bilayer as a function of the coupling ratio $\lambda=J/J'$. Results are shown for various linear system
  sizes $L$. The temperature was chosen $\beta J'=2L$, low enough to
be in the scaling regime. The cutoff $\Lambda=8L^3$ was chosen large enough to cover the coupling range $J/J'\lesssim1$.  The dynamical critical exponent of this model
 is $z=1$ and $\eta=0.034$.}
\label{fig:bilayer}
\end{figure}

We consider as an example the {\it quantum phase transition}
in a bilayer Heisenberg antiferromagnet whose ground state changes from 
quantum disordered to N\'eel ordered as the ratio
$\lambda=J/J'$ of intra-plane (J) to inter-plane ($J'$) coupling is increased \cite{bilayer}. 
From the histograms generated within {\it one}
simulation we can calculate the staggered structure factor $S(\pi,\pi)$ of the system
at {\it any value} of $\lambda$. 
In Fig. \ref{fig:bilayer}   we show a scaling plot
of $S(\pi,\pi) / L^{2-z-\eta}$ as a function of $\lambda$.
In short simulations, taking only a few days on an 800 MHz Pentium-III CPU,
 we find the quantum critical point at $\lambda=0.396$, which again
compares well with earlier results \cite{bilayer}.

To summarize, we have presented Monte Carlo algorithms for the direct
calculation of the free energy of a quantum system, based on a stochastic
series expansion representation of the partition function. Our
algorithms employ a variant of the Wang-Landau sampling to achieve a flat
histogram and provide the free energy as well as thermodynamic averages
accurately over a whole temperature or coupling range. The algorithms can
be used with any of the update schemes developed for the SSE
representation and require only minor modifications to existing programs.
Parallelization of the algorithms can be done like in the classical case by splitting the $n$-range into
multiple random walks over a shorter range.

Our algorithms are efficient not only at second order phase transitions but also at first order ones, where standard local and cluster update methods fail. The algorithms open up new possibilities for quantum Monte Carlo simulations: similar to the classical case we expect the algorithms to be efficient also for disordered systems and work is in progress to apply the methods to quantum spin glasses. Also, the ability to calculate the free energy and entropy will be useful for investigations of entropy-driven phase transitions, such as the reentrant melting transition observed in bosonic systems and anisotropic quantum magnets \cite{schmid}.

We thank D.P. Landau for discussions and acknowledge support of the Swiss
National Science Foundation.

\end{document}